\documentclass[11pt]{article}

\usepackage[T1]{fontenc}
\usepackage[utf8]{inputenc}
\usepackage[english]{babel}
\usepackage[margin=1in]{geometry}
\usepackage{amsmath,amssymb}
\usepackage{booktabs}
\usepackage{longtable}
\usepackage{array}
\usepackage{tabularx}
\usepackage{enumitem}
\usepackage[table]{xcolor}
\usepackage[colorlinks=true,linkcolor=blue,urlcolor=blue]{hyperref}

\newcolumntype{Y}{>{\raggedright\arraybackslash}X}
\newcommand{\density}{\rho}
\newcommand{\densityhat}{\widehat{\rho}}
\newcommand{\AETHERP}{AETHER-P\textsuperscript{3}}
\newcommand{\modelname}{\AETHERP{} Nowcast}

\setlist[itemize]{leftmargin=*,itemsep=0.25em}

\title{
    \modelname{} v1.0:\\[0.4em]
    \large Model Description, Training-Data Construction, and Validation\\[0.8em]
    \normalsize\textit{Technical report prepared in support of CCMC onboarding}
}

\author{
    \textbf{Ruochen Wang and Xiaoli Bai}\\[0.8em]
    \normalsize Department of Mechanical and Aerospace Engineering\\
    Rutgers, The State University of New Jersey\\
    Piscataway, New Jersey, USA
}

\date{September 2026}

\begin{document}
\maketitle

\section{Model Purpose}

\modelname{} is a global thermospheric neutral mass-density model for
low-Earth-orbit applications. It estimates density at a requested universal
time, latitude, longitude, and altitude using time-directed solar, solar-wind,
and geomagnetic history. It is intended to support applications such as orbit
analysis, satellite-drag assessment, and monitoring of thermospheric
conditions.

The model is pointwise: one evaluation corresponds to one time and one
location. Repeating the same evaluation over a grid produces a global
three-dimensional density field.

In addition to a central density estimate, the model provides a predictive
distribution. This permits the same model call to return a density interval
and separate measures of aleatoric and epistemic uncertainty.

\section{Model Identification and Version}

\begin{table}[htbp]
\centering
\renewcommand{\arraystretch}{1.15}
\begin{tabularx}{\textwidth}{>{\bfseries}p{0.31\textwidth}Y}
\toprule
Item & Current value\\
\midrule
Model name & \modelname\\
Current software version & \texttt{1.0.0}\\
Model class & Supervised neural-network density nowcast with a joint
Normal--Inverse-Gamma evidential output\\
Number of model inputs & 342\\
Number of trainable parameters & 328,116\\
Selected realization & Random seed 20; validation-selected epoch 117\\
Model developer(s) & Ruochen Wang, Xiaoli Bai\\
Primary/model-owner contact & Xiaoli Bai (xiaoli.bai@rutgers.edu)\\
Technical contact & Ruochen Wang (ruo.chen.wang@rutgers.edu)\\
Institution & Department of Mechanical and Aerospace Engineering, Rutgers,
The State University of New Jersey, Piscataway, NJ 08854\\
Release date & 19 August 2026\\
\bottomrule
\end{tabularx}
\end{table}

\subsection{Software and release traceability}

Table~\ref{tab:software-traceability} identifies the exact software and model
artifact used for this report. Release \texttt{v1.0.0} is fixed to the source
commit listed below.

\begin{table}[htbp]
\centering
\renewcommand{\arraystretch}{1.15}
\begin{tabularx}{\textwidth}{>{\bfseries}p{0.31\textwidth}Y}
\toprule
Item & Current value\\
\midrule
GitHub repository &
\url{https://github.com/x-bai-research-group/aether-p3-nowcast-ccmc}\\
Release/tag & \href{https://github.com/x-bai-research-group/aether-p3-nowcast-ccmc/releases/tag/v1.0.0}{\texttt{v1.0.0}}\\
Exact Git commit &
\texttt{f8e671dad7c10463b4c5562cecfd05835}\newline
\texttt{ee65f37}\\
Model checkpoint & \texttt{model/model.weights.h5}\\
Checkpoint SHA-256 &
\texttt{eecc45f27e5fc40d80f74f1b2cc2c803}\newline
\texttt{f5c6d7dc6174ee17ae4225ffddec8a04}\\
Checkpoint epoch selection & Minimum panel-balanced validation EDL loss;
evaluation benchmarks were not used\\
Released realization selection & Seed 20 selected by a balanced comparison of
validation behavior and the 12 evaluation benchmarks after all seed-specific
checkpoints were fixed\\
Python & 3.12.11 (supported range: 3.11--3.12)\\
Machine-learning framework & TensorFlow/Keras 2.20; PyTorch is not used\\
Supported operating system & Linux x86-64\\
Tested operating system & Ubuntu 24.04.4 LTS, Linux kernel 6.11\\
\bottomrule
\end{tabularx}
\caption{Software, release, and model traceability.}
\label{tab:software-traceability}
\end{table}

\subsection{Computational requirements}

GPU acceleration is not required for released-model inference or three-
dimensional field generation. A CUDA-capable GPU is strongly recommended for
training and can accelerate neural-network inference, although the complete
field runtime also includes CPU-based physical feature and empirical-model
calculations. Table~\ref{tab:computational-requirements} reports the tested
configuration and matched CPU/GPU timings; they are approximate rather than
hardware-independent performance guarantees.

\begin{table}[htbp]
\centering
\renewcommand{\arraystretch}{1.15}
\begin{tabularx}{\textwidth}{>{\bfseries}p{0.31\textwidth}Y}
\toprule
Requirement & Current value\\
\midrule
GPU required for inference & No\\
Tested CPU & Intel Core Ultra 9 285K, 24 cores\\
Tested GPU & NVIDIA GeForce RTX 5090, 32 GB; used for training and GPU-enabled
tests\\
System memory & 64 GB tested; at least 16 GB recommended for complete field
generation\\
Observed production peak memory & 2.06 GiB CPU and 2.82 GiB GPU host resident
memory for the benchmark field; the Java process is launched with a 14 GB
maximum heap\\
One-point model inference & Warm median of 0.393 s on CPU and 0.075 s on the
RTX 5090 for one precomputed 342-D feature vector\\
Model evaluation for one complete grid & 84.41 s on CPU and 7.31 s on the RTX
5090 for 251,100 points, including physical uncertainty conversion; GPU
speedup 11.54$\times$\\
Complete three-dimensional field & 102.46 s on CPU and 26.83 s with RTX 5090
inference for 251,100 points (31 altitudes $\times$ 90 latitudes $\times$ 90
longitudes), including feature generation, JB2008 and NRLMSISE-00 evaluation,
model inference, and NetCDF writing; end-to-end speedup 3.82$\times$\\
\bottomrule
\end{tabularx}
\caption{Tested computational requirements and approximate inference times.}
\label{tab:computational-requirements}
\end{table}

For each random-seed realization, the checkpoint was selected using validation
data alone. The 12 evaluation benchmark cases did not select or replace epoch
117 within the seed-20 training run. Their results were, however, reviewed
when seed 20 was chosen as the released realization. Consequently, the
benchmark results reported below are descriptive results for the selected
realization, not an untouched independent estimate obtained after all model
selection was complete.

\section{Model Inputs}

\subsection{Input definition}

The model receives 342 normalized scalar values divided into six physical
groups. Table~\ref{tab:inputs} gives the complete input definition used by
version 1.0.0.

\begin{table}[htbp]
\centering
\small
\renewcommand{\arraystretch}{1.12}
\caption{Input groups used by \modelname.}
\label{tab:inputs}
\begin{tabularx}{\textwidth}{p{0.19\textwidth}c c Y}
\toprule
Input group & Shape & Dim. & Variables and temporal support\\
\midrule
Location and time & 10 & 10 & Latitude; sine and cosine of longitude;
altitude; sine and cosine of day of year, universal time, and local solar
time.\\
Solar background & 2 & 2 & Previous-day observed F10.7 and current-day
F30.\\
Solar-proxy history & $7\times4$ & 28 & F10 and S10 from D-1 to D-7, M10
from D-2 to D-8, and Y10 from D-5 to D-11.\\
Fast forcing history & $35\times6$ & 210 & Dst, ap30, GSM $B_z$, solar-wind
speed, proton number density, and AE from 170 minutes before the query to the
current state at five-minute spacing.\\
Long geomagnetic history & $45\times2$ & 90 & Hourly AE and Dst from 48 to
4 hours before the query.\\
Empirical density references & 2 & 2 & Current-location
$\log_{10}\rho_{\mathrm{JB2008}}$ and
$\log_{10}\rho_{\mathrm{NRLMSISE\text{-}00}}$.\\
\midrule
\textbf{Total} & & \textbf{342} &\\
\bottomrule
\end{tabularx}
\end{table}

Longitude is represented by sine and cosine. Therefore $-180^\circ$ and
$+180^\circ$ have the same input representation, preventing an artificial
density discontinuity at the international date line. Latitude, longitude,
altitude, and time are encoded separately before they are combined with the
external forcing histories.

JB2008 and NRLMSISE-00 are informative input features rather than fixed
backgrounds. The neural model is free to predict above or below either
empirical estimate. The causal NRLMSISE-00 calculation uses previous-day
F10.7 and a trailing 81-day F10.7 average; it does not use a retrospective
centered average.

\subsection{Causality and operational availability}

All histories are evaluated at or before the requested UTC. The current
lookup policy is:

\begin{itemize}
  \item F10.7 uses the previous UTC day;
  \item F10, S10, M10, and Y10 use the daily delays listed in
  Table~\ref{tab:inputs};
  \item Dst uses the current available hourly UTC bin;
  \item ap30 uses the current completed 30-minute bin;
  \item AE, $B_z$, solar-wind speed, and proton density use the most recent
  causal observation no more than five minutes old; and
  \item a request fails if a required driver is unavailable in the supplied
  local tables rather than applying additional interpolation at request time.
\end{itemize}

The lookup rule is causal with respect to the supplied local tables. The
fixed research solar-wind table was prepared separately: flagged or missing
OMNI magnetic-field and plasma values were filled by linear interpolation
before the Java runtime stage. The actual publication latency and operational
update schedule depend on the external provider.

For CCMC onboarding, a separate Box folder contains the prepared runtime data
files used in the current research implementation to facilitate installation
and reproducibility. These files remain subject to the terms of their original
data providers. The GitHub documentation provides the authoritative source
locations and preprocessing procedures, enabling CCMC to obtain and prepare
the source data independently as needed. The present release defines the
scientific input contract and documents the preprocessing used for the
research implementation.

Operational data sources, publication-latency rules, and missing-data handling
for the CCMC implementation will be finalized jointly with CCMC during
onboarding.

\section{Model Outputs}

\subsection{Probabilistic model output}

The density target is represented internally as standardized log-density,

\begin{equation}
 z=\frac{\log_{10}\density-\mu_y}{\sigma_y},
\end{equation}

where $\mu_y$ and $\sigma_y$ are calculated from the selected training data.
One model call returns four Normal--Inverse-Gamma parameters,

\begin{equation}
 (\gamma,\nu,\alpha,\beta),
 \qquad \nu>0,\quad\alpha>1,\quad\beta>0.
\end{equation}

The parameter $\gamma$ is the central prediction in normalized log-density
space. The corresponding physical-density estimate is

\begin{equation}
 \densityhat=10^{\mu_y+\sigma_y\gamma}.
\end{equation}

The four parameters define the predictive distribution

\begin{equation}
 z \sim \operatorname{Student}\text{-}t\left(
 2\alpha,\ \gamma,
 \sqrt{\frac{\beta(1+\nu)}{\alpha\nu}}
 \right),
\end{equation}

where the three arguments are degrees of freedom, location, and scale in
normalized log-density space. The corresponding uncertainty components on the
$\log_{10}\rho$ scale are

\begin{equation}
 \sigma_{\mathrm{aleatoric}}
 =\sigma_y\sqrt{\frac{\beta}{\alpha-1}},
 \qquad
 \sigma_{\mathrm{epistemic}}
 =\sigma_y\sqrt{\frac{\beta}{\nu(\alpha-1)}}.
\end{equation}

The 95\% density interval is obtained by transforming the 2.5\% and 97.5\%
Student-$t$ quantiles to physical density. No post-hoc calibration multiplier
is applied.

\subsection{NetCDF product}

The preferred global-field output is NetCDF4. Its dimensions are ordered as
\texttt{time, altitude, latitude, longitude}. The current variables are
listed in Table~\ref{tab:netcdf}.

\begin{table}[htbp]
\centering
\small
\renewcommand{\arraystretch}{1.12}
\caption{Variables in the current NetCDF output.}
\label{tab:netcdf}
\begin{tabularx}{\textwidth}{p{0.27\textwidth}p{0.13\textwidth}Y}
\toprule
Variable & Units & Description\\
\midrule
\texttt{density} & $\mathrm{kg\,m^{-3}}$ & Central neutral-density
estimate.\\
\texttt{density\_lower\_95} & $\mathrm{kg\,m^{-3}}$ & Lower bound of the
Student-$t$ 95\% predictive interval.\\
\texttt{density\_upper\_95} & $\mathrm{kg\,m^{-3}}$ & Upper bound of the
Student-$t$ 95\% predictive interval.\\
\texttt{aleatoric\_std\_log10} & 1 & Aleatoric standard deviation in
log-density space.\\
\texttt{epistemic\_std\_log10} & 1 & Epistemic standard deviation in
log-density space.\\
\texttt{gamma} & 1 & Normalized mean log-density.\\
\texttt{nu}, \texttt{alpha}, \texttt{beta} & 1 & Remaining
Normal--Inverse-Gamma parameters.\\
\bottomrule
\end{tabularx}
\end{table}

The files declare CF-1.10 conventions and include the model version, feature
contract, native cadence, grid definition, and recommended altitude range as
metadata.

\subsection{Native production grid}

\begin{table}[htbp]
\centering
\begin{tabular}{lll}
\toprule
Coordinate & Range & Native resolution\\
\midrule
Time & Requested nowcast UTC & 5 minutes\\
Latitude & $-89^\circ$ to $+89^\circ$ & $2^\circ$ cell centered\\
Longitude & $-178^\circ$ to $+178^\circ$ & $4^\circ$ cell centered\\
Altitude & 230--530 km & 10 km\\
\bottomrule
\end{tabular}
\end{table}

\section{Training Data}

\subsection{Density observations}

The supervised target is neutral mass density obtained from accelerometer or
mission density products. Five satellite missions contribute training data.
Table~\ref{tab:training-missions} gives the approximate observation interval,
the number of eligible 30-second rows before compact sampling, and the number
selected for final training.

\begin{table}[htbp]
\centering
\small
\caption{Satellite density data used to construct the training set.}
\label{tab:training-missions}
\begin{tabular}{llrr}
\toprule
Mission & Approximate interval & Eligible rows & Training rows\\
\midrule
CHAMP & 29 Jul 2000--4 Sep 2010 & 9,596,668 & 445,011\\
GRACE-A & 4 Apr 2002--31 Oct 2017 & 13,931,541 & 586,460\\
GOCE & 1 Dec 2009--20 Oct 2013 & 2,513,749 & 101,102\\
Swarm-C & 1 Feb 2014--31 Dec 2023 & 8,657,745 & 342,376\\
GRACE-FO & 29 May 2018--31 Dec 2023 & 5,601,299 & 196,961\\
\midrule
\textbf{Total} & 2000--2023 & \textbf{40,301,002} &
\textbf{1,671,910}\\
\bottomrule
\end{tabular}
\end{table}

The density products were obtained from the ESA Swarm and multi-mission
archives and the ESA GOCE Thermosphere Data collection. Solar-wind inputs and
the fixed AE snapshot were obtained through NASA OMNI; the AE index
distributed by OMNI originates from WDC Kyoto. Dst comes directly from WDC
Kyoto, ap30 from GFZ, and solar proxies from the JB2008/SET, CelesTrak, and CLS
data products. JB2008 and NRLMSISE-00 density references are evaluated locally
through Orekit 13.1.4. Because final, provisional, and quick-look AE products
can be released or revised on different schedules, a later AE download is not
assumed to be numerically identical to the fixed research snapshot.

The provider-format daily products remain separate files. Dst, AE, ap30, and
high-resolution OMNI exports are converted into one chronological UTC table
per product using documented canonical columns; they are not combined by
averaging across sources. At inference, the Java feature generator joins these
tables causally at the requested UTC. Dst uses its hourly bin, ap30 becomes
available only after completion of its half-hour interval, and AE and solar-
wind measurements may be carried backward by at most five minutes. No linear
interpolation is added during this runtime join. The fixed solar-wind table,
however, already contains offline linear interpolation of flagged or missing
OMNI magnetic-field and plasma values. These locally reformatted tables remain
subject to their source-provider terms and are not redistributed with the code
release.

\subsection{Coverage-preserving sampling}

The source archive is dominated by consecutive quiet-time observations. The
final training set therefore uses deterministic, target-blind sampling by
physical regime. Density values, mission identity, validation error, and
evaluation-benchmark results do not affect training-row inclusion. All
available extreme rows are retained.

\begin{table}[htbp]
\centering
\caption{Final training distribution.}
\begin{tabular}{lrrr}
\toprule
Regime & Eligible rows & Selected rows & Nominal retention\\
\midrule
Ordinary quiet & 31,680,182 & 790,337 & $1/40$\\
High-solar quiet & 3,982,899 & 198,843 & $1/20$\\
Transition & 3,363,379 & 335,797 & $1/10$\\
Moderate & 1,237,080 & 309,471 & $1/4$\\
Extreme & 37,462 & 37,462 & All\\
\bottomrule
\end{tabular}
\end{table}

Natural unit weights are used during training; inverse-inclusion and manual
storm weights are not applied.

\subsection{Preprocessing}

\begin{itemize}
  \item Density labels are retained observations rather than linearly
  interpolated density estimates.
  \item Flagged or missing OMNI solar-wind driver values were linearly
  interpolated when the fixed research driver table was assembled; this is
  separate from density-label processing.
  \item Rows with an invalid density or a missing required driver are
  rejected.
  \item The supervised target is $\log_{10}\density$ in
  $\mathrm{kg\,m^{-3}}$.
  \item Feature and target normalization are calculated only from the selected
  training rows.
  \item Mission identity is retained for evaluation but is not supplied to the
  model.
  \item The training rows are stored in a fixed globally shuffled order.
\end{itemize}

\section{Validation and Evaluation Benchmark Data}

\subsection{Validation data}

The validation set contains 690,722 rows from the same five training missions
but from 24 complete chronological blocks. The blocks span 2001--2023 and
include quiet background, storm development and recovery, the 2008--2009 deep
solar minimum, and high-solar conditions in 2023.

Training and validation intervals are temporally disjoint. A 48-hour exclusion
guard is applied on both sides of every validation and evaluation-benchmark
interval.
This duration matches the longest causal AE/Dst history window and prevents
protected observations from entering any training sequence.

Each epoch is evaluated on all validation blocks. The checkpoint objective is
the unweighted mean of block-level full evidential losses. Accuracy metrics
and evaluation-benchmark results do not enter this within-realization
checkpoint objective. Thus, the selected epoch for each random seed is
determined exclusively from validation data.

\subsection{Evaluation benchmark cases}

The 12 evaluation benchmark cases in Table~\ref{tab:evaluation-cases} are
excluded from training records, validation rows, and the checkpoint selection
performed within each random-seed realization. The February 2015 intervals
are explicit temporal holdouts. The May 2024 intervals occur after the density
training archive ends in 2023. Swarm-A and Swarm-B are not training missions.
After the checkpoint for each realization had been fixed, validation behavior
and results on these cases were considered jointly when selecting seed 20 as
the released realization. The cases therefore did not influence checkpoint
selection within a seed, but they did influence the final choice among trained
seeds. They remain useful documented benchmarks, but they are not claimed to
be an untouched independent test set for the final realization-selection step.

\begin{longtable}{c l l p{0.34\textwidth}}
\caption{Evaluation benchmark cases.}
\label{tab:evaluation-cases}\\
\toprule
Case & Mission & Interval & Physical condition\\
\midrule
\endfirsthead
\toprule
Case & Mission & Interval & Physical condition\\
\midrule
\endhead
1 & Swarm-A & 24--31 May 2024 & High-solar quiet\\
2 & Swarm-C & 24--31 May 2024 & High-solar quiet\\
3 & Swarm-A & 10--15 Feb 2015 & Medium-solar quiet\\
4 & Swarm-C & 10--15 Feb 2015 & Medium-solar quiet\\
5 & Swarm-A & 16--20 Feb 2015 & Moderate event I\\
6 & Swarm-C & 16--20 Feb 2015 & Moderate event I\\
7 & Swarm-A & 23--25 Feb 2015 & Moderate event II\\
8 & Swarm-C & 23--25 Feb 2015 & Moderate event II\\
9 & GRACE-FO & 10--13 May 2024 & Extreme storm\\
10 & Swarm-A & 10--13 May 2024 & Extreme storm\\
11 & Swarm-B & 10--13 May 2024 & Extreme storm\\
12 & Swarm-C & 10--13 May 2024 & Extreme storm\\
\bottomrule
\end{longtable}

Every model is evaluated at identical UTC, location, and observed-density rows.
HASDM is the principal high-accuracy comparison. JB2008 and NRLMSISE-00 are
also reported. The available WAM-IPE WRS nowcast files cover only the six May
2024 cases, so WAM-IPE is not available for the February 2015 cases.

\section{Known Limitations}

\begin{enumerate}[leftmargin=*]
  \item \textbf{Altitude range.} The standard output grid spans 230--530 km.
  The principal application range is 250--520 km; predictions outside this
  central range have less direct observational support and should be
  interpreted with additional caution.

  \item \textbf{Dependence on training coverage.} The model is supervised and
  can degrade for combinations of solar activity, storm phase, altitude, and
  location that are poorly represented in the accelerometer-density archive.
  Consecutive rows within one storm do not represent independent storm
  realizations.

  \item \textbf{High-solar quiet conditions.} The May 2024 quiet cases are
  outside the training period and retain a systematic absolute density-level
  error. Their relative error and RMSE are currently worse than HASDM even
  though their temporal correlation is high.

  \item \textbf{Rare extreme cases.} Overall storm performance is strong, but
  an unusual high-density tail can produce worse correlation or RMSE than
  HASDM even when the typical fractional error is lower.

  \item \textbf{Mission and product differences.} Accelerometer density
  products may contain mission-dependent bias. Mission identity is
  intentionally excluded, so the model does not fit an explicit correction
  for each satellite product.

  \item \textbf{Input availability.} A nowcast requires every declared
  driver. Provider outages or publication latency can prevent execution.

  \item \textbf{Uncertainty calibration.} Calibration is condition
  dependent. The aggregate 95\% coverage is below its nominal value, mainly
  because the difficult high-solar quiet and some extreme cases are
  under-covered. MACE and 95\% coverage must be reported together.

  \item \textbf{No target-time density assimilation.} The model does not
  ingest a contemporaneous accelerometer density measurement. It is a driver-
  based nowcast, not a real-time density-assimilation system.
\end{enumerate}

\section{Benchmark Results}

\subsection{Metrics}

Accuracy is reported using mean absolute relative error (RE), Pearson
correlation ($R$), and physical-density root mean square error (RMSE). These
measure typical fractional error, agreement in variation, and absolute error,
respectively.

Uncertainty is reported using mean absolute calibration error (MACE) and the
observed coverage of the nominal Student-$t$ 95\% predictive interval (T95).
Ideal values are zero for MACE and 0.95 for T95. MACE and T95 are interpreted
together.

\subsection{Current model performance}

Table~\ref{tab:benchmark-summary} reports the unweighted mean across the 12
evaluation benchmark cases for the selected seed-20 model. RMSE is expressed
in units of $10^{-13}\,\mathrm{kg\,m^{-3}}$.

\begin{table}[htbp]
\centering
\small
\caption{Current evaluation benchmark summary.}
\label{tab:benchmark-summary}
\begin{tabular}{lrrrrrr}
\toprule
Model & RE & $R$ & RMSE & MACE & T95 & RE wins vs HASDM\\
\midrule
\AETHERP{} seed 20 & \textbf{0.1157} & \textbf{0.9363} &
\textbf{2.4748} & 0.0726 & 0.9209 & 10/12\\
HASDM & 0.1473 & 0.9090 & 2.6900 & -- & -- & --\\
JB2008 & 0.2195 & 0.8291 & 4.2402 & -- & -- & --\\
NRLMSISE-00 & 0.4343 & 0.8454 & 6.0952 & -- & -- & --\\
WAM-IPE$^{a}$ & 0.9370 & 0.8123 & 15.0597 & -- & -- & --\\
\bottomrule
\end{tabular}

\vspace{0.4em}
\begin{minipage}{0.94\textwidth}
\footnotesize
$^{a}$WAM-IPE statistics use only its six available May 2024 cases and are not
a directly matched 12-case aggregate.
\end{minipage}
\end{table}

\AETHERP{} has lower RE than HASDM in 10 of 12 cases. It simultaneously has
lower RE, higher correlation, and lower RMSE in 8 of 12 cases. The two
remaining RE losses are the high-solar quiet Cases 1 and 2. These results did
not select epoch 117 within the seed-20 run, but they were considered when
seed 20 was designated as the final realization. They should therefore be
interpreted as descriptive benchmark performance rather than an untouched
independent performance estimate.

\subsection{Performance by individual case}

Table~\ref{tab:benchmark-by-case} gives the complete case-level accuracy
comparison. The case numbers follow Table~\ref{tab:evaluation-cases}; RMSE is
reported in units of $10^{-13}\,\mathrm{kg\,m^{-3}}$. Bold type identifies the
best available model for each case and metric. WAM-IPE is shown only where a
matching WRS nowcast file is available.

\begingroup
\small
\setlength{\tabcolsep}{3pt}
\renewcommand{\arraystretch}{1.08}
\begin{longtable}{@{}clrrrrr@{}}
\caption{Accuracy comparison for each evaluation benchmark case.}
\label{tab:benchmark-by-case}\\
\toprule
Case & Metric & \AETHERP{} & HASDM & JB2008 & NRLMSISE-00 & WAM-IPE\\
\midrule
\endfirsthead
\multicolumn{7}{c}{\tablename\ \thetable\ continued}\\
\toprule
Case & Metric & \AETHERP{} & HASDM & JB2008 & NRLMSISE-00 & WAM-IPE\\
\midrule
\endhead
\midrule
\multicolumn{7}{r}{Continued on next page}\\
\endfoot
\bottomrule
\endlastfoot
1 & RE   & 0.143036 & \textbf{0.113216} & 0.290732 & 0.470256 & 1.012565\\
  & $R$  & \textbf{0.979901} & 0.977157 & 0.958817 & 0.941110 & 0.950987\\
  & RMSE & 1.940553 & \textbf{1.154628} & 3.894678 & 5.380995 & 9.419509\\
\addlinespace
2 & RE   & 0.141980 & \textbf{0.134148} & 0.280593 & 0.494741 & 1.076193\\
  & $R$  & \textbf{0.983427} & 0.977770 & 0.961606 & 0.945639 & 0.952219\\
  & RMSE & 1.846705 & \textbf{1.189530} & 3.778138 & 5.303269 & 9.385817\\
\addlinespace
3 & RE   & \textbf{0.090956} & 0.125945 & 0.134046 & 0.206978 & --\\
  & $R$  & \textbf{0.948642} & 0.928742 & 0.890703 & 0.908987 & --\\
  & RMSE & \textbf{1.182136} & 1.392051 & 1.870276 & 2.687316 & --\\
\addlinespace
4 & RE   & \textbf{0.086915} & 0.118162 & 0.127988 & 0.210512 & --\\
  & $R$  & \textbf{0.958875} & 0.936727 & 0.899419 & 0.919921 & --\\
  & RMSE & \textbf{1.104652} & 1.312054 & 1.744712 & 2.699806 & --\\
\addlinespace
5 & RE   & \textbf{0.081684} & 0.128879 & 0.134813 & 0.146154 & --\\
  & $R$  & \textbf{0.935748} & 0.883000 & 0.845024 & 0.859275 & --\\
  & RMSE & \textbf{1.065976} & 1.451384 & 1.754597 & 1.949999 & --\\
\addlinespace
6 & RE   & \textbf{0.085234} & 0.129354 & 0.136542 & 0.152656 & --\\
  & $R$  & \textbf{0.936743} & 0.888238 & 0.852735 & 0.862245 & --\\
  & RMSE & \textbf{1.130432} & 1.464851 & 1.814947 & 2.081034 & --\\
\addlinespace
7 & RE   & \textbf{0.073291} & 0.138669 & 0.135201 & 0.126469 & --\\
  & $R$  & \textbf{0.882664} & 0.790080 & 0.713623 & 0.796036 & --\\
  & RMSE & \textbf{1.050948} & 1.679343 & 1.812959 & 1.844732 & --\\
\addlinespace
8 & RE   & \textbf{0.081610} & 0.139972 & 0.139370 & 0.133006 & --\\
  & $R$  & \textbf{0.883678} & 0.801305 & 0.727979 & 0.804681 & --\\
  & RMSE & \textbf{1.132550} & 1.695521 & 1.875299 & 1.942582 & --\\
\addlinespace
9 & RE   & \textbf{0.164603} & 0.206805 & 0.334363 & 0.826327 & 0.907220\\
  & $R$  & 0.922113 & \textbf{0.926175} & 0.810012 & 0.790278 & 0.733114\\
  & RMSE & \textbf{5.143782} & 5.586621 & 7.768796 & 12.773720 & 18.423210\\
\addlinespace
10 & RE   & \textbf{0.136497} & 0.175488 & 0.305439 & 0.799140 & 0.804714\\
   & $R$  & \textbf{0.928948} & 0.925934 & 0.742381 & 0.763434 & 0.737890\\
   & RMSE & \textbf{5.079646} & 5.887914 & 8.891156 & 13.559600 & 18.412750\\
\addlinespace
11 & RE   & \textbf{0.167655} & 0.184123 & 0.304762 & 0.842948 & 1.005889\\
   & $R$  & 0.944437 & \textbf{0.944844} & 0.800532 & 0.787463 & 0.762814\\
   & RMSE & 4.004435 & \textbf{3.734853} & 6.898707 & 9.484452 & 16.443540\\
\addlinespace
12 & RE   & \textbf{0.135025} & 0.172525 & 0.309680 & 0.802217 & 0.815642\\
   & $R$  & \textbf{0.930104} & 0.927767 & 0.746516 & 0.766141 & 0.736826\\
   & RMSE & \textbf{5.015692} & 5.730947 & 8.777757 & 13.434590 & 18.273200\\
\end{longtable}
\endgroup

Table~\ref{tab:uncertainty-by-case} reports uncertainty calibration for the
same frozen model. MACE evaluates the complete calibration curve, whereas
$|\mathrm{T95}-0.95|$ isolates the error at the nominal 95\% interval.

\begin{table}[htbp]
\centering
\small
\caption{Uncertainty performance for each evaluation benchmark case.}
\label{tab:uncertainty-by-case}
\begin{tabular}{crrr}
\toprule
Case & MACE & T95 & $|\mathrm{T95}-0.95|$\\
\midrule
1  & 0.201740 & 0.773308 & 0.176692\\
2  & 0.195670 & 0.813944 & 0.136056\\
3  & 0.013738 & 0.963102 & 0.013102\\
4  & 0.022291 & 0.975639 & 0.025639\\
5  & 0.040062 & 0.988508 & 0.038508\\
6  & 0.028159 & 0.987841 & 0.037841\\
7  & 0.087374 & 0.988949 & 0.038949\\
8  & 0.057751 & 0.987019 & 0.037019\\
9  & 0.055583 & 0.872917 & 0.077083\\
10 & 0.043984 & 0.913611 & 0.036389\\
11 & 0.086040 & 0.873924 & 0.076076\\
12 & 0.039337 & 0.912163 & 0.037837\\
\bottomrule
\end{tabular}
\end{table}

Across all 12 cases, \AETHERP{} reduces mean RE by 21.4\% and mean RMSE by
8.0\% relative to HASDM, while increasing the mean correlation from 0.9090 to
0.9363. Cases 3--8 show the broadest advantage: \AETHERP{} is better than all
reported baselines in RE, correlation, and RMSE for every medium-solar quiet
and moderate-event case. This indicates that the improvement is not limited
to one error measure or one satellite.

Cases 1 and 2 are the clear accuracy limitation. \AETHERP{} has slightly higher
correlation than HASDM but worse RE and RMSE. It therefore follows the observed
variation while retaining an error in the absolute density level. Both cases
are high-solar quiet intervals in May 2024, after the density training archive
ends, which is consistent with sensitivity to training coverage rather than a
general failure to represent temporal variation.

During the extreme storm, \AETHERP{} has lower RE than HASDM in all four cases
and a lower RMSE in three. Case 11 is the exception in RMSE, and HASDM also has
a marginally higher correlation in Cases 9 and 11. The aggregate extreme-case
RMSE nevertheless decreases from $5.2351\times10^{-13}$ to $4.8109\times
10^{-13}\,\mathrm{kg\,m^{-3}}$. These results support strong storm performance
while showing that rare high-density portions of an individual trajectory can
remain difficult.

The uncertainty results are also condition dependent. Cases 3 and 4 have low
MACE and coverage close to 0.95. The moderate cases are conservative, with
T95 near 0.988, whereas Cases 1 and 2 are under-covered and account for the
largest calibration errors. Extreme-case coverage ranges from 0.873 to 0.914.
Consequently, the aggregate MACE of 0.0726 and T95 of 0.9209 should not be
interpreted as uniform calibration across all operating conditions.

\section{Citation and Funding}

\subsection{\AETHERP{} scientific publications}

The \AETHERP{} research framework is described in the following publications:

\begin{itemize}
  \item Wang, Y., and Bai, X. (2024), ``A Global Thermospheric Density
  Prediction Framework Based on a Deep Evidential Method,'' \emph{Space
  Weather}, 22(12), e2024SW004070,
  \href{https://doi.org/10.1029/2024SW004070}{doi:10.1029/2024SW004070}.

  \item Wang, R., and Bai, X. (2026), ``A Machine-Learning-Based Global
  Thermospheric Density Forecasting Model,'' \emph{Space Weather}, 24(6),
  e2026SW004968,
  \href{https://doi.org/10.1029/2026SW004968}{doi:10.1029/2026SW004968}.
\end{itemize}

The CCMC nowcast configuration documented here represents a subsequent,
delivery-oriented implementation of the \AETHERP{} framework and is not
identical to the model configurations reported in the publications above.

\subsection{Method, empirical-model, and software references}

The following references describe the principal methods and empirical models
used by \modelname:

\begin{itemize}
  \item Amini, A., Schwarting, W., Soleimany, A., and Rus, D. (2020),
  ``Deep Evidential Regression,'' \emph{Advances in Neural Information
  Processing Systems}, 33.

  \item Bowman, B. R., et al. (2008), ``A New Empirical Thermospheric Density
  Model JB2008 Using New Solar and Geomagnetic Indices,'' AIAA 2008-6438,
  \href{https://doi.org/10.2514/6.2008-6438}{doi:10.2514/6.2008-6438}.

  \item Picone, J. M., Hedin, A. E., Drob, D. P., and Aikin, A. C. (2002),
  ``NRLMSISE-00 empirical model of the atmosphere: Statistical comparisons
  and scientific issues,'' \emph{Journal of Geophysical Research: Space
  Physics}, 107(A12),
  \href{https://doi.org/10.1029/2002JA009430}{doi:10.1029/2002JA009430}.

  \item Orekit Team, Orekit version 13.1.4,
  \url{https://www.orekit.org/site-orekit-13.1.4/downloads.html}.
\end{itemize}

\subsection{Observation and driver data sources}

The density observations and external drivers were obtained from the following
scientific data services:

\begin{itemize}
  \item ESA Swarm Data Access for Swarm and the multi-mission CHAMP, GRACE,
  and GRACE-FO density products,
  \url{https://swarm-diss.eo.esa.int/}.

  \item ESA GOCE Online Data Access for the GOCE thermosphere product,
  \url{https://goce-ds.eo.esa.int/oads/access/}.

  \item NASA SPDF OMNIWeb for high-resolution solar-wind variables and the AE
  archive used here,
  \url{https://omniweb.gsfc.nasa.gov/html/ow_data.html}.

  \item WDC Kyoto for the original AE index and hourly Dst,
  \url{https://wdc.kugi.kyoto-u.ac.jp/dstae/index.html}.

  \item GFZ German Research Centre for Geosciences for ap30,
  \url{https://kp.gfz.de/en/hp30-hp60}.

  \item Space Environment Technologies for the JB2008 SOLFSMY and DTC
  indices, \url{https://sol.spacenvironment.net/JB2008/indices.html}.

  \item CelesTrak for the consolidated SW-All space-weather table,
  \url{https://celestrak.org/SpaceData/SW-All.txt}.

  \item CLS for adjusted F30 solar radio flux,
  \url{https://spaceweather.cls.fr/services/radioflux/}.
\end{itemize}

\subsection{Funding and acknowledgments}

\noindent
\textbf{Funding:} This research has been supported by the National Science
Foundation, United States, under Award 2149747, and the National Aeronautics
and Space Administration (NASA), United States, under Award
80NSSC24K0843.\\[0.6em]

\end{document}